 \documentclass[aps,prl,twocolumn,showpacs,superscriptaddress,longbibliography]{revtex4-2}
\usepackage{graphicx} 
\usepackage{amsmath}
\usepackage{graphicx,epstopdf}
\usepackage{gensymb}
\usepackage{multibib}
\newcommand{\qv}{\mathbf q}

\newcommand{\be}{\begin{equation}}
\newcommand{\ee}{\end{equation}}
\newcommand{\bea}{\begin{eqnarray}}
\newcommand{\eea}{\end{eqnarray}}
\newcommand{\bse}{\begin{subequations}}
	\newcommand{\ese}{\end{subequations}}

\usepackage{color}
\usepackage[colorlinks,bookmarks=false,citecolor=darkblue,linkcolor=red,urlcolor=blue]{hyperref}

\definecolor{darkred}{rgb}{0.7,0.0,0.0}

\definecolor{darkblue}{rgb}{0,0.02,0.45}
\def\cdbl{\color{darkblue}}
\definecolor{darkgreen}{rgb}{0.02,0.45,0.0}

\definecolor{violet}{rgb}{0.8,0.2,0.6}

\begin{document}

\title{Intermediate field-induced phase of the honeycomb magnet BaCo$_2$(AsO$_4$)$_2$}

\author{Prashanta K. Mukharjee}
\email{pkmukharjee92@gmail.com}
\affiliation{Experimental Physics VI, Center for Electronic Correlations and Magnetism, Institute of Physics, University of Augsburg, 86159 Augsburg, Germany}

\author{Bin Shen}
\affiliation{Experimental Physics VI, Center for Electronic Correlations and Magnetism, Institute of Physics, University of Augsburg, 86159 Augsburg, Germany}

\author{Sebastian Erdmann}
\affiliation{Experimental Physics VI, Center for Electronic Correlations and Magnetism, Institute of Physics, University of Augsburg, 86159 Augsburg, Germany}

\author{Anton Jesche}
\affiliation{Experimental Physics VI, Center for Electronic Correlations and Magnetism, Institute of Physics, University of Augsburg, 86159 Augsburg, Germany}

\author{Julian Kaiser}
\affiliation{Experimental Physics VI, Center for Electronic Correlations and Magnetism, Institute of Physics, University of Augsburg, 86159 Augsburg, Germany}

\author{Priya~R.~Baral}
\affiliation{Laboratory for Neutron Scattering and Imaging (LNS), Paul Scherrer Institute (PSI), CH-5232, Villigen, Switzerland}

\author{Oksana Zaharko}
\affiliation{Laboratory for Neutron Scattering and Imaging (LNS), Paul Scherrer Institute (PSI), CH-5232, Villigen, Switzerland}

\author{Philipp Gegenwart}
\email{gegenwart@physik.uni-augsburg.de}
\affiliation{Experimental Physics VI, Center for Electronic Correlations and Magnetism, Institute of Physics, University of Augsburg, 86159 Augsburg, Germany}

\author{Alexander A. Tsirlin}
\email{altsirlin@gmail.com}
\affiliation{Felix Bloch Institute for Solid-State Physics, University of Leipzig, 04103 Leipzig, Germany}

\date{\today}

\begin{abstract}
	We use magnetometry, calorimetry, and high-resolution capacitive dilatometry, as well as single-crystal neutron diffraction to explore temperature-field phase diagram of the anisotropic honeycomb magnet BaCo$_2$(AsO$_4)_2$. Our data reveal four distinct ordered states observed for in-plane magnetic fields. Of particular interest is the narrow region between 0.51 and 0.54\,T that separates the up-up-down order from the fully polarized state and coincides with the field range where signatures of the spin-liquid behavior have been reported. We show that magnetic Bragg peaks persist in this intermediate phase, thus ruling out its spin-liquid nature. However, the simultaneous nonmonotonic evolution of nuclear Bragg peaks suggests the involvement of the lattice, witnessed also in other regions of the phase diagram where large changes in the sample length are observed upon entering the magnetically ordered states. Our data highlight the importance of lattice effects in BaCo$_2$(AsO$_4)_2$.
\end{abstract}

\maketitle

{\cdbl\textit{Introduction.}}
Quantum spin liquid (QSL) is an exotic state of matter in which strong quantum fluctuations prevent magnetic long-range order (LRO) down to very low temperatures~\cite{Broholm2020}. The exactly solvable $S =\frac{1}{2}$ honeycomb Kitaev model possessing bond-dependent nearest-neighbor Ising-type interactions offers a promising venue for stabilizing the QSL ground state with the excitation spectrum described by emergent Majorana fermions and gauge fluxes~\cite{Kitaev2}. Experimental realization of this model has been of significant interest. The $d^5$ (Ir$^{4+}$, Ru$^{3+}$)~\cite{Khaliullin017205} and, more recently, $d^7$ (Co$^{2+}$)~\cite{Liu014407} transition-metal ions were proposed as suitable building blocks of the Kitaev magnets. 

The majority of Ir-, Ru-, and Co-based honeycomb materials show long-range order in zero magnetic field~\cite{Winter2017,Kim2022}. However, it was conjectured that they may lie in the vicinity of the QSL phase, and external field can be used to suppress long-range order, thus giving way to the spin liquid. Indeed, magnetic field applied along a suitable direction leads to a rapid suppression of the ordered phase in $\alpha$-RuCl$_3$ and $\beta$-Li$_2$IrO$_3$ above the critical fields of $B_c\simeq 7$\,T and 2.8\,T, respectively~\cite{Takagi2019,Ruiz2017,Majumder2019}. The behavior of $\alpha$-RuCl$_3$ above $B_c$ remains vividly debated. Whereas quantized thermal Hall effect~\cite{Yokoi2021,Lefrancois2022,Bruin2022}, oscillations in thermal conductivity~\cite{Czajka2021,Lefrancois2023,Bruin2022b}, and a distinct excitation continuum~\cite{Banerjee2018,Sahasrabudhe2003} observed in this field range might be vestiges of the spin-liquid behavior, no distinct phase associated with this putative spin liquid could be identified~\cite{Bachus2020,Schoenemann2020,Gass2020,Bachus2021}.

Here, we focus on the Co-based honeycomb magnet, BaCo$_2$(AsO$_4)_2$ (BCAO) recently proposed as a Kitaev candidate~\cite{Zhong6953} with several intriguing analogies to $\alpha$-RuCl$_3$. The two materials show quite different sequences of the magnetically ordered states, double-zigzag-like and field-induced up-up-down (\textit{uud}) orders in BCAO~\cite{Regnault660,Regnault1979,Regnaulte00507} vs. different types of zigzag order in $\alpha$-RuCl$_3$~\cite{Balz2021}. Moreover, a detailed study of spin dynamics~\cite{Halloran2023}, along with \textit{ab initio} calculations~\cite{Maksimov2022}, suggested that an easy-plane $J_1-J_3$ Hamiltonian is more suitable for BCAO than an extended Kitaev model commonly accepted for $\alpha$-RuCl$_3$~\cite{Winter2017}. On the other hand, both materials show several apparent similarities too. They can be polarized by moderate in-plane fields and demonstrate an unusual behavior right above $B_c$. 
An excitation continuum in the 
paramagnetic state~\cite{Zhang01403} and a small linear contribution to the thermal conductivity near $B_c$~\cite{Tu2023}, which is often 
associated with spinon excitations, are reminiscent of the spin-liquid 
physics with the possible field-induced spin-liquid phase.

In the following, we scrutinize the behavior of BCAO in applied magnetic fields.
Using several thermodynamic probes, we map out the temperature-field phase diagram of this material and uncover the previously overlooked intermediate phase that appears near $B_c$ and separates the \textit{uud} order from the fully polarized state. This intermediate phase coincides with the field range where linear-in-temperature thermal conductivity has been reported. However, it does not show characteristic signatures of a spin liquid, namely, the absence of the long-range magnetic order, because the magnetic Bragg peak persists in this phase. Our results rule out the formation of the field-induced spin-liquid phase in BCAO and further highlight the importance of spin-lattice coupling in this material, as evidenced by
large changes in the sample length across field-induced phase transitions.

\begin{figure}
	\includegraphics [width = \linewidth]{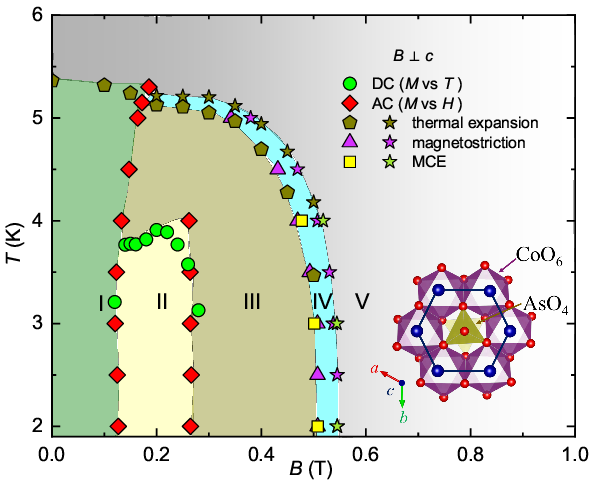}
	\caption{Magnetic field-temperature ($H-T$) phase diagram of BaCo$_2$(AsO$_4$)$_2$ for $B \perp c$ constructed from magnetization $M(T,B)$, dilatometry $L(T,B)$ and MCE measurements. Phase diagrams for different in-plane field directions are shown in Fig.~S12. Inset: A honeycomb motif containing CoO$_6$ octahedra and AsO$_4$ tetrahedra at the center. The individual Co-honeycomb layers are stacked along the $c$-axis with Ba atoms separating them.}
	\label{Fig1}
\end{figure} 

{\cdbl\textit{Methods.}}
Dark-pink-colored single crystals of BCAO were synthesized by the flux method~\cite{supplement}. High quality of these crystals was confirmed by an extensive mapping with Laue diffraction and by neutron diffraction~(see Fig.~S1). Temperature- and field-dependent magnetization was measured using the commercial SQUID magnetometer (MPMS3, Quantum Design), with orientation-dependent magnetization data collected using the sample rotation stage of MPMS3. Specific heat was measured using a standard relaxation technique in Quantum Design Physical Properties Measurement System (QD - PPMS). The magnetic Gr\"uneisen parameter ($\Gamma_m$), which equals the adiabatic magnetocaloric effect (MCE), was determined by the alternating-field method~\cite{Tokiwa2011}. Thermal expansion was measured with the aid of a compact ultrahigh-resolution capacitive dilatometer in the QD PPMS~\cite{Kuchler083903}. The length change as a function of temperature or field was recorded using a capacitance bridge (Andeen Haggerling 2550A). Neutron diffraction data from a single crystal were collected on the ZEBRA diffractometer at SINQ, Paul Scherrer Institute. 

{\cdbl\textit{Phase diagram.}}
Previous report revealed an almost isotropic behavior of BCAO within the honeycomb plane~\cite{Halloran2023}, in contrast to $\alpha$-RuCl$_3$ that showed different phase boundaries for two nonequivalent in-plane field directions, $b$ and $b^*$~\cite{Bachus2021,Balz2021}. Our data for BCAO confirm this isotropic in-plane behavior. Therefore, in Fig.~\ref{Fig1} we merged the data points obtained for different field directions, whereas individual phase diagrams for $B\|b$ and $B\|b^*$ can be found in the Supplemental Material~[Fig.~S12]. In magnetometry, we determine temperature and field values of the transitions from the peak positions of the respective derivatives. In dilatometry, we consider the peak positions of linear thermal expansion coefficient ($\alpha$) and magnetostriction ($\lambda$), while in the MCE ($\Gamma_m$) data, we identify the transitions from the positions of the minima and zero-crossings~\cite{Bachus2021}. 

Two magnetically ordered states of BCAO, the incommensurate double-zigzag order and commensurate \textit{uud} order~\cite{Regnault660,Regnault1979,Regnaulte00507}, are labeled as phases I and III in our phase diagram, respectively. Phase V is the fully polarized state. Additionally, we uncover phases II and IV that separate I from III and III from V, respectively. 
Phase IV is of particular interest, as it coincides with the field range of the putative spin liquid in BaCo$_2$(AsO$_4)_2$~\cite{Tu2023}.

\begin{figure}
	\includegraphics[width = \linewidth]{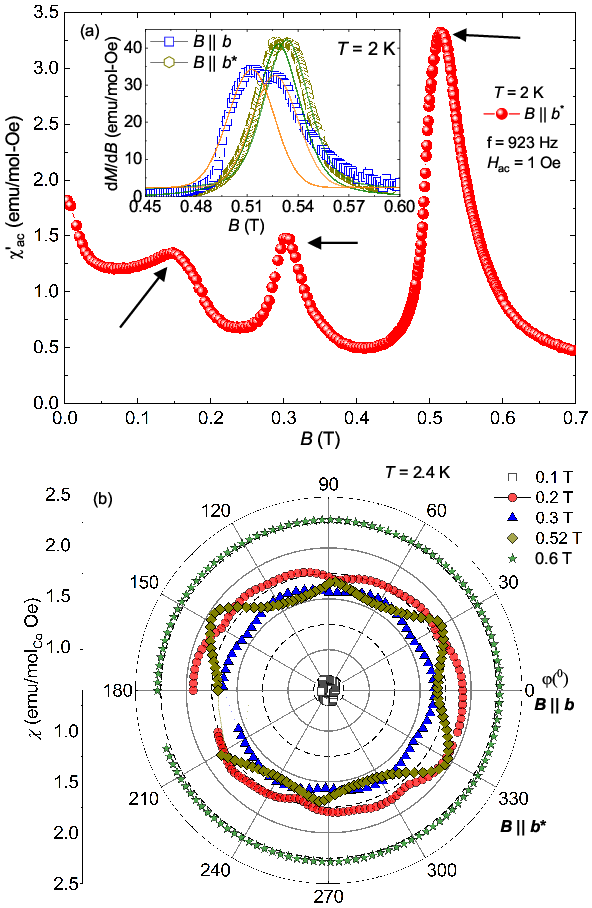}
	\caption{\label{Fig2}(a) Magnetic field dependence of the ac-susceptibility $\chi_{ac}^{\prime}$ at 2~K for $B || b^*$. The arrow marks indicate the position of the respective phase boundaries. The inset shows the differential susceptibility $dM/dB$ obtained from dc-magnetization measurements. The double peak in $dM/dB$ confirms the formation of phase IV. The solid lines are visual guides, illustrating the separation of the two peaks. (b) Angle-dependent magnetic susceptibility measured at $T=2.4$~K for different magnetic fields while rotating the crystal in the $ab$ plane. Note the change in the positions of the minima and maxima between 0.3\,T (phase III) and 0.52\,T (phase IV).}
\end{figure}

Temperature-dependent dc-magnetization of BCAO~[Fig.~S2] shows an abrupt drop at $T_N$ followed by a broad maximum that appears between 0.12 and 0.26\,T, the field range that we identify as phase II. The boundaries of this phase are most clearly visible in the ac-susceptibility that shows two consecutive peaks as a function of field (Fig.~\ref{Fig2}a). The lower boundary of phase II is characterized by the abrupt increase in $M(B)$, suggesting that the transformation from the antiferromagnetic double-zigzag into the partially polarized \textit{uud} state starts upon entering phase II. This transformation is first-order in nature, as witnessed by the large field hysteresis~[Fig.~S4]. The anomalies associated with phase II disappear around 4\,K, whereas phases I and III survive up to $5.2-5.3$\,K. 

The transition between I and III involves a flip of ferromagnetic zigazag spin chains from the $\uparrow\uparrow\downarrow\downarrow\uparrow\uparrow\downarrow\downarrow$ into the $\uparrow\uparrow\downarrow\uparrow\uparrow\downarrow$ configuration, which is essentially the creation and shift of domain walls. It is plausible that such a shift requires thermal energy and at lower temperatures involves an intermediate region, which we identify as phase II. This phenomenology may be similar to the isostructural BaCo$_2$(PO$_4)_2$ where magnetization curve shows distinct steps that correspond to discrete shifts of the domain walls, albeit at temperatures below 1\,K only~\cite{Wang2023}.

\begin{figure}
	\includegraphics[width = \linewidth]{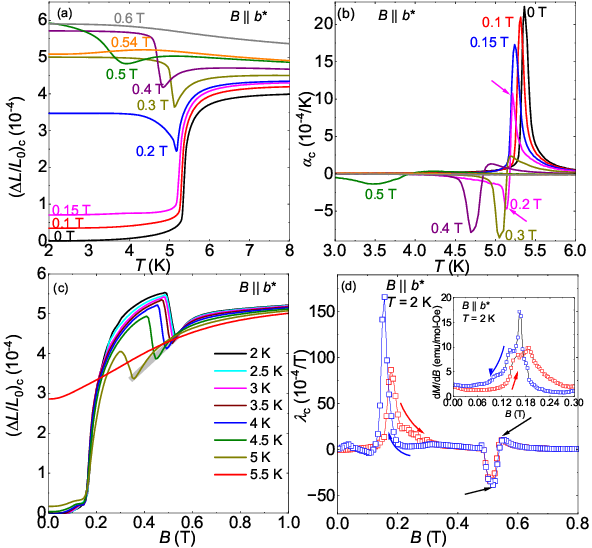}
	\caption{\label{Fig3} Thermal expansion and magnetostriction plots of BCAO. (a) Normalized relative length changes vs temperature measured while warming at various fields. The curves are shifted vertically for clarity. (b) Plot of the linear TE coefficient ($\alpha$) in different fields. For $B = 0$~T, in phase I, $\alpha$ shows a positive peak. As we move into phase II, $\alpha$ bifurcates to two peaks with opposite sign as indicated by the arrows. (c) Normalized relative length changes vs magnetic field measured while sweeping magnetic field up at several temperatures above and below $T_N$. The light grey line is a guide to eye showing the systematic shift of minimum around $B_c$. (d) Plot of linear MS coefficient $\lambda$ vs. $B$ at $T = 2$~K. Two prominent peaks are indicated by arrows around $B_c$. The transition around 0.15~T exhibits significant hysteresis, consistent with the field-dependent differential susceptibility [inset], reflecting its first-order nature.}
\end{figure}

We now turn to thermal expansion (TE) and magnetostriction (MS) where all phase transitions are clearly visible thanks to the large lattice effects involved. BCAO shrinks along $c$ upon entering phase I and expands along $c$ upon entering phase III, but the most interesting behavior is seen in temperature-dependent thermal expansion measured above 0.2\,T where sample length changes non-monotonically indicating two consecutive phase transitions, first with the decrease and then with the increase in $c$ upon cooling (Fig.~\ref{Fig3}a). A similar behavior is seen in field-dependent sample length that changes non-monotonically on going from phase III into phase V (Fig.~\ref{Fig3}c). The two transitions can be tracked using the minima and maxima in linear thermal expansion coefficient $\alpha = (1/L_0)(d\Delta L/dT)$ and magnetostriction $\lambda = (1/L_0)(d \Delta L/dB)$ (Fig.~\ref{Fig3}b,d), resulting in the distinct region of phase IV that envelops phase III and separates it from phase V. This phase IV is remarkably different from phase II because it extends in temperature all the way up to 5.3\,K where BCAO enters its paramagnetic state. Moreover, the transitions associated with phase IV should be second-order or weakly first-order in nature, as they do not show any significant hysteresis.

{\cdbl\textit{Intermediate phase.}}
Phase IV is most clearly seen in the thermal expansion and magnetostriction data. It is further evidenced by the differential susceptibility $dM/dB$ where the double peak at $0.50-0.54$\,T demonstrates that phase III transforms into phase V via the intermediate phase IV (Fig.~\ref{Fig2}a, inset). The MCE measurement $\Gamma_m$ displays the minimum and zero-crossing at the III--IV and IV--V boundaries, respectively~[see Fig.~S10]. Finally, angle-dependent susceptibility reveals a clear change from the 6-fold symmetric response at 0.3\,T, as expected for the \textit{uud} state (phase III) in the $R\bar 3$ crystal structure, to a weakly deformed hexagon at 0.52\,T within phase IV (Fig.~\ref{Fig2}b). This hexagon is turned by $30^{\circ}$, such that the minima and maxima of the susceptibility are swapped, indicating a qualitative change in the in-plane magnetic anisotropy between phases III and IV.

\begin{figure}
	\includegraphics[width = \linewidth]{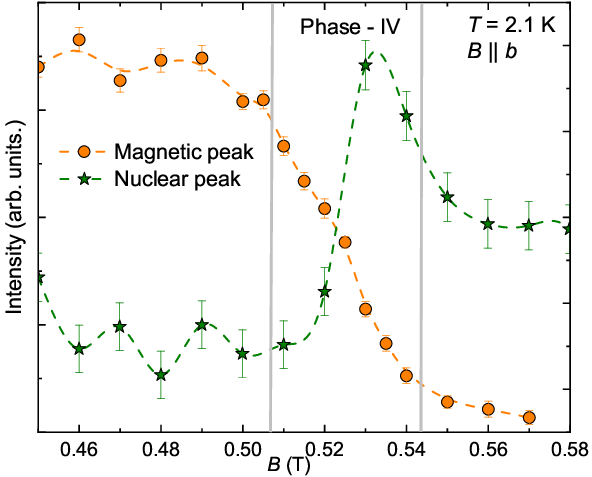}
	\caption{\label{Fig4} Magnetic field dependence of 
		the magnetic Bragg peak at $\mathbf q=(1/3,0,-4/3)$ and nuclear Bragg 
		peak at $\mathbf q=(3,0,0)$ across phase IV for $B\|b$.}
\end{figure}

We further probed phase IV using neutron diffraction. In phase III, we observed the magnetic Bragg peak at $\qv=(\frac13,0,-\frac43)$ in agreement with the previous studies~\cite{Regnault1979,Halloran2023}. This peak is also observed between 0.50 and 0.54\,T, in the field range of phase IV, but gradually loses its intensity and vanishes when the fully polarized state is reached [see Fig.~\ref{Fig4}]. In order to probe whether this evolution of the magnetic Bragg peak could merely indicate a gradual reduction in the ordered magnetic moment, we further tracked the nuclear Bragg peaks across the same field range. A direct transformation between phases III and V with the phase coexistence between 0.50 and 0.54\,T should lead to the gradual increase in the intensity of the nuclear Bragg peaks through the shift of the magnetic intensity from $\qv=(\frac13,0,-\frac43)$ to $\qv=0$ as magnetization increases. However, the $\qv=0$ intensities change non-monotonically, as shown in Fig.~\ref{Fig4} for a representative nuclear Bragg peak. We thus conclude that a distinct phase IV appears between 0.50 and 0.54\,T, in agreement with thermodynamic measurements where the double transition is observed in differential susceptibility, thermal expansion, and magnetostriction.

{\cdbl\textit{Pressure dependence of the transitions.}}
Magnetic transitions in BCAO are accompanied by remarkably large lattice effects. For example, the zero-field transition between the paramagnetic phase and phase I corresponds to the relative length change of $4\times 10^{-4}$, about four times larger than in $\alpha$-RuCl$_3$~\cite{Gass2020}. It indicates the unusually strong spin-lattice coupling in BCAO.

To explore manifestations of this spin-lattice coupling, we analyze the initial pressure dependence of the magnetic transitions using the peaks in $\alpha$ and $\lambda$. According to Maxwell's relation, $\lambda V=-(dM/dp)_{T, B}$ where $V$ is the volume and $M$ is the magnetization. Whereas we do not measure volume magnetostriction in our experiments, the qualitative behavior given by the positive ($\lambda_{c} > 0$) and negative peaks ($\lambda_{c} < 0$) in linear magnetostriction suggests that external pressure should decrease the magnetization in phase I and increase it in phase III. The former trend is indeed confirmed by our isothermal magnetization measurements performed under hydrostatic pressure~[Fig.~S11c].

Pressure dependence of the transition temperature, $dT_N/dp$, is given by the Clausius-Clapeyron and Ehrenfest relations for the first- and second-order transitions, respectively. From the maxima and minima in $\alpha_c$, we expect $T_N$ to increase with pressure in phase I and decrease with pressure in phase III. These trends are again consistent with the direct magnetization measurements under hydrostatic pressure using both our data~\cite{supplement} and the earlier report~\cite{Huyan184431}. 
Phase diagram of BCAO drastically changes under pressure because the $\frac13$-plateau is no longer seen at 1.6\,GPa~\cite{supplement}. It means that phase III gives way to some other phase that allows a gradual increase in the magnetization, in contrast to the \textit{uud} order that keeps the magnetization fixed at $\frac13$ of the saturated value. One interesting possibility is that phase III may be fully replaced by phase IV under pressure. 

{\cdbl\textit{Discussion.}}
Our data rule out the spin-liquid behavior of BCAO in the vicinity of $B_c$. Although we observe a distinct pre-saturation phase, the persistence of the magnetic Bragg peak indicates the presence of long-range order all the way up to 0.54\,T where the fully polarized state sets in. This absence of the field-induced spin-liquid phase undermines the consideration of BCAO as a candidate Kitaev material.

At first glance, phase IV of BCAO could be paralleled to the so-called zz2 phase reported in $\alpha$-RuCl$_3$~\cite{Balz2021,Bachus2021} above 6\,T. This phase is magnetically ordered and characterized by the swapped maxima and minima of the angle-dependent susceptibility~\cite{Balz2021}, similar to phase IV of BCAO (Fig.~\ref{Fig2}b). The zz2 phase of $\alpha$-RuCl$_3$ further shows the six-layer periodicity of the magnetic order, in contrast to the three-layer periodicity in lower fields. However, we did not detect any significant change in the interlayer component of the magnetic order of BCAO within phase IV. Therefore, the direct analogy to $\alpha$-RuCl$_3$, even on the level of its magnetically ordered phases, seems unlikely.

From the theoretical perspective, possible phases of $J_1-J_3$ frustrated honeycomb ferromagnets have been studied in detail~\cite{Jiang2023,Bose2023,Watanabe2023}. Whereas spin liquid could be stable in a very narrow range of parameters~\cite{Bose2023}, the observation of the magnetic Bragg peak up to the full saturation at 0.54\,T speaks against the formation of this spin-liquid phase in BCAO. The V-type order proposed as one of the possible field-induced phases~\cite{Maksimov2023} is also unlikely, because it should be characterized by a different propagation vector compared to the \textit{uud} state. We are thus led to conclude that the existing theoretical framework does not offer a microscopic explanation for the formation of phase IV. It seems likely that the appearance of this phase is influenced by the spin-lattice coupling that has not been considered in Kitaev materials to date, although it is known to stabilize nontrivial field-induced phases in pyrochlore~\cite{Miyata2020,Gen2023} and triangular~\cite{Nakajima2013} magnets. Experimental signatures of the spin-lattice coupling in BCAO include: i) large lattice changes upon the magnetic transitions, and ii) loss of the 6-fold symmetry within phase IV, in contrast to phase III characterized by the 6-fold-symmetric magnetic susceptibility.


The intermediate phase observed in our work does not cover the full field range where indications of the spin-liquid have been reported. Indeed, the linear-in-temperature contribution to the thermal conductivity extends up to 0.6\,T~\cite{Tu2023}, well into the fully polarized state. However, the onset of this unusual behavior at 0.50\,T is clearly correlated with the formation of phase IV and not with reaching the fully polarized state around 0.55\,T, so the presence of this intermediate phase must be taken into account when the microscopic scenario of the spin-liquid-like behavior of BCAO is looked for.

{\cdbl\textit{Conclusions.}}
Our comprehensive study of the anisotropic honeycomb magnet BaCo$_2$(AsO$_4$)$_2$ rules out the existence of a distinct spin-liquid phase near $B_c$ and reveals instead a narrow intermediate phase with persistent magnetic order. Large lattice changes accompanying all magnetic transitions, as well as the loss of the 6-fold symmetry within the intermediate phase, indicate the importance of spin-lattice coupling in this material. Whereas BaCo$_2$(AsO$_4$)$_2$ does not show any clear-cut manifestations of the Kitaev physics, it may serve as an interesting case of a frustrated honeycomb ferromagnet where competing magnetic phases are controlled by lattice effects.

\acknowledgements
We thank Pavel Maksimov, Sasha Chernyshev, Lukas Janssen, and Radu Coldea for fruitful discussions. P.K.M was supported by the Alexander von Humboldt foundation. The neutron single cyrstal diffraction experiment was performed at SINQ, Paul Scherrer Institute, Villigen, Switzerland. This work was funded by the Deutsche Forschungsgemeinschaft (DFG, German Research Foundation) -- TRR 360 -- 492547816 (subproject B1).\\

Experimental data associated with this manuscript are available from \cite{Mukharjee_BCAO1}.\\

\noindent \textcolor{red}{$^*$}pkmukharjee92@gmail.com\\
\textcolor{red}{$^\dagger$}gegenwart@physik.uni-augsburg.de\\
\textcolor{red}{$^\ddag$}altsirlin@gmail.com
\bibliography{reff_BCAO}
\widetext
\clearpage
\begin{center}
	\textbf{\large Supplementary Material for \\ Intermediate field-induced phase of the honeycomb magnet BaCo$_2$(AsO$_4$)$_2$}
\end{center}

\setcounter{equation}{0}
\setcounter{figure}{0}
\setcounter{table}{0}
\setcounter{page}{1}
\makeatletter
\setcounter{section}{0}
\renewcommand{\thesection}{S-\Roman{section}}
\renewcommand{\thetable}{S\arabic{table}}
\renewcommand{\theequation}{S\arabic{equation}}
\renewcommand{\thefigure}{S\arabic{figure}}
\renewcommand{\bibnumfmt}[1]{[S#1]}
\renewcommand{\citenumfont}[1]{S#1}

\section{Crystal Growth}
\begin{figure}[htb]
	\includegraphics[width= 14 cm]{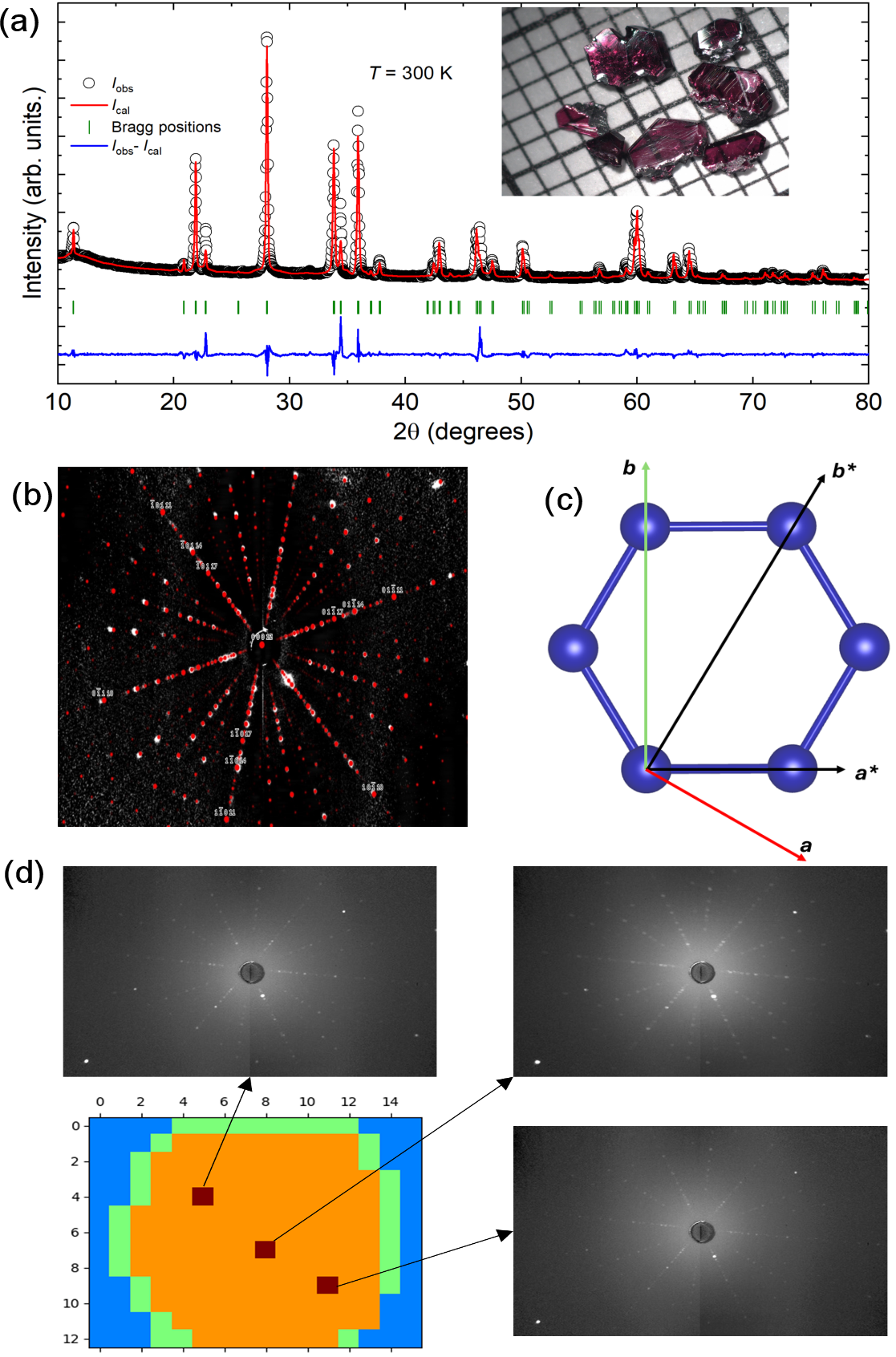}
	\caption{(a) Powder x-ray diffraction data for BaCo$_2$(AsO$_4$)$_2$ collected at $T = 300$~K (black circles). The red lines indicate the Rietveld refinement fit of the data. The Bragg peaks are denoted by green vertical bars and the bottom blue line denotes the difference between the experimental and calculated intensities. Inset: Dark pink crystals on a millimeter paper. (b) Laue pattern of a crystal taken from the $c$ - direction. The red spots represent simulated Laue patterns generated using Crystal Maker software. (c) A Schematic sketch of the convention used for different crystallographic orientations. (d) Laue back-scattering grid analysis was utilized to map out possible domains in the single crystal of BCAO. The diffraction images were color-coded: orange for good diffraction images, green for weak patterns (edges), and blue for no observable patterns (sample holder). It was confirmed that different spots on the crystal surface had the same crystallographic orientation, thus excluding multi-domain growth.
		\label{FigS1}} 
\end{figure}

Single crystals of BaCo$_2$(AsO$_4$)$_2$ were grown by a similar method as reported in \cite{Zhong6953}. Dark-pink, hexagonal plate-like crystals (inset of Fig.~\ref{FigS1}a) were separated from the flux by washing in hot water. A few crystals were crushed into powders, and room-temperature powder XRD data were collected. To determine the sample orientation and possible stacking disorder, we collected X-ray Laue back reflection patterns (Photonic Sciences) (Fig.~\ref{FigS1}b). The clear Laue spots indicate the high quality of the BCAO single crystal. In contrast to findings in other honeycomb materials~\cite{Yan2019}, our Laue pattern clearly shows the absence of rod-like diffuse spots, effectively ruling out the presence of defects/stacking faults in BCAO. Moreover, through extensive Laue photography covering the entirety of the crystal surface, we have confirmed that BCAO crystals exhibit a monodomain structure, contrasting with the multi-domain structure, which is often observed in $\alpha$-RuCl$_3$~\cite{Johnson2015}.

A few crystals were crushed into fine powder, and room-temperature x-ray diffraction data were recorded using the Empyrean powder diffractometer from PANalytical (CuK$_{\alpha}$ radiation, $\lambda_{avg} \simeq 1.5418~\text{Å}$). Rietveld refinement for the collected data performed using the "FULLPROF" software~\cite{Carvajal55} confirms the single phase of BCAO and returns lattice constants of $a = 5.0052(2)~\text{Å}$ and $c = 23.481(1)~\text{Å}$, which are consistent with the previous report~\cite{Zhong6953}.

\section{Magnetization and Specific Heat}
\begin{figure}[htb]
	\includegraphics[width= 14 cm]{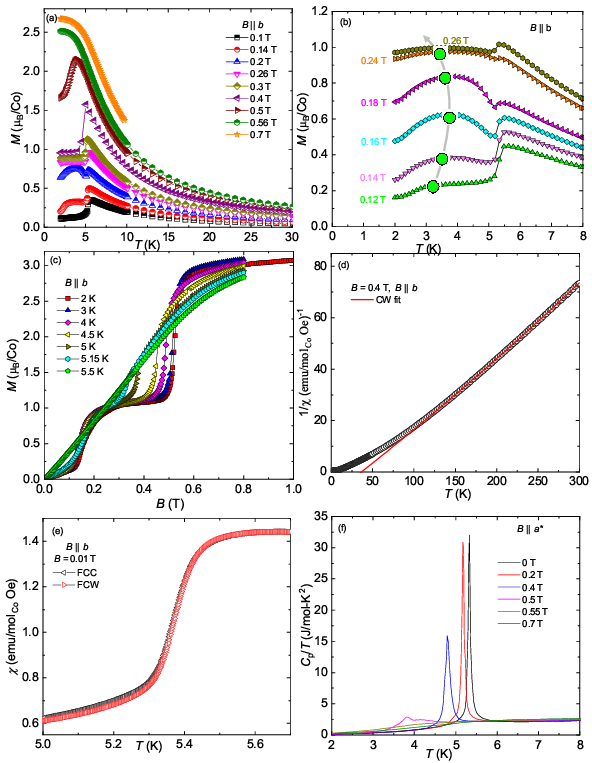}
	\caption{(a) Magnetic susceptibility vs. temperature under magnetic fields up to 0.7~T. (b) $\chi(T)$ measured for 0.12~T $\leq B \leq$ 0.26~T where a broad hump has been observed. The green circles indicate the corresponding broad maximum positions. (c) Magnetic isotherms measured for 5.5~K $\leq T \leq 2$~K. (d) Inverse susceptibility vs. temperature along with the CW fit (red solid line). (d) Inverse susceptibility vs. temperature along with CW fit (red solid line). (e) $\chi(T)$ measured at $B = 0.01~$T for FCC-FCW cycle. (f) Specific-heat divided by temperature [$C_{p}/T$] vs. $T$ for 0~T $\leq H \leq$ 0.7~T.
		\label{FigS2}}
\end{figure}

The evolution of $T_N$ was tracked from the magnetic susceptibility measured in different applied magnetic fields for the $B \parallel b$ orientation as shown in Fig.~\ref{FigS2}(a). In Fig.~\ref{FigS2}(b), we show a broad hump around $3.2$~K, observed in the field range of 0.12 to 0.26~T and associated with phase II in the phase diagram. The transition at $T_N \simeq 5.35$~K decreases systematically before completely vanishing at $\sim 0.55$~T. Magnetic isotherms collected for 5.5~K $\leq T \leq 2~$K are shown in Fig.~\ref{FigS2}(c). To estimate the magnetic couplings between the Co$^{2+}$ ions, we have performed Curie-Weiss (CW) fit of the susceptibility data using  $\chi = \chi_0 + C/(T-\theta)$, where $\chi_0$ is the temperature-independent contribution, $C$ is the Curie constant, $\theta$ is the CW temperature. The fit in the region $150-300$~K yields a positive CW temperature $\theta_{\parallel, b} \simeq 37$~K suggesting predominant ferromagnetic  interactions. From the obtained value of $C$, the effective paramagnetic moment ($\mu_{\rm eff})$ is estimated to be $\mu_{\rm eff \parallel, b} \simeq 5.58 \mu_B$ [see Fig.~\ref{FigS2}(d)]. These estimations are in agreement with the literature~\cite{Zhong6953,Tu2023}. The deviation of the obtained effective moment (taking $= g \sqrt{J(J+1)}$, where $g$ is the Landé $g$-factor and $J = 1/2$) from the theoretical value for spin-$\frac12$ ($1.73~\mu_B$) arises due to the significant spin-orbit coupling, which is common in Co$^{2+}$ quantum magnets. 

The second-order nature of the transition into phase I is supported by examining the $T$-dependent susceptibility (at $0.01$~T) throughout field-cooled-cooling (FCC) and field-cooled-warming (FCW) processes [see Fig.~\ref{FigS2}(e)]. The curves for both directions are reversible with no hysteresis. We have also measured temperature-dependent specific heat for $B \parallel a^{*}$ as shown in Fig.~\ref{FigS2}(f). In zero-field, $C_p(T)$ shows a sharp peak at $T_N$, which gradually decreases with increasing magnetic field and vanishes above 0.5~T, in line with other experiments.

\section{Dilatometry}
High-resolution capacitive dilatometry was used to measure length changes $L(T,H)$. The experimental design is described in Ref.~\cite{Kuchler083903}. In all our experiments, we measured the relative length changes along the $c$-direction, while for field-dependent measurements, the magnetic field was oriented along the $b$ and $b^{*}$-direction. The normalized relative length change is defined as follows,
\begin{equation}
	\frac{\Delta L (T,B)}{L_0} = \frac{L(T,B)- L (300~K,0~T)}{L (300~K,0~T)}.
\end{equation}
Here, $L_0$ is the sample thickness at 300~K and 0~T. Throughout our analysis, $\Delta L(T,B)$ values are standardized to the value at 2~K and 0~T. For thermal expansion, the data were collected during cooling-warming cycles with the sweep rate of 0.2~K/min. Magnetostriction measurements were performed up to 1~T with the sweep rate of 60~mT/min. To calculate the pressure dependence of $T_N$, we use the Ehrenfest relation for a second-order transition, 
\begin{equation}
	\frac{dT_N}{dp} = V_mT_N \frac{\Delta \beta}{\Delta C_{p}}
\end{equation}
where $V_m$ = 2.54 $\times 10^{-5}$~m$^{3}$/mol is the molar volume,  $\Delta \beta = \Delta \alpha_{a^{*}} + \Delta \alpha_{b} + \Delta \alpha_{c}$ is the jump in the volume thermal expansion coefficient at $T_N$, and $\Delta C_{p}$ is the jump in $C_{p}$ at $T_N$. For this purpose a crystal was cut into a rectangular shape and corresponding thermal expansion and specific heat are measured as shown in Fig.~\ref{FigS7}.

Figure~\ref{FigS9}(a) depicts the magnetostriction and field-dependent differential susceptibility ($dM/dB$) together for $T = 2$~K. The first-order transformation (phase I $\leftrightarrow$ phase II) is evidenced by a sharp peak both in $\lambda_{c}$ and $dM/dB$ plots. Near $B_c$ (in phase IV),  $\lambda_{c}$ shows two peaks, whereas $dM/dB$ features a broad behavior. Figure~\ref{FigS9}(b) shows the variation of $\lambda_c$ with magnetic fields for 5.5~K $\leq T \leq 2$~K.

\section{Magnetocaloric Effect}
Field dependence of the magnetic Gr\"uneisen parameter ($\Gamma_m$) is used to track field-induced states around $B_c$. This is done via the magnetocaloric effect (MCE) measurement under quasiadiabatic conditions. The magnetic Gr\"uneisen parameter is defined as
\begin{equation}
	\Gamma_m = \frac{1}{T} \frac{\partial T}{\partial B}.
\end{equation}
This method utilizes the alternating-field technique~\cite{Tokiwa2011}, wherein a weakly oscillating magnetic field with an amplitude ($\Delta B$) is applied. This induces oscillations in the sample temperature ($\Delta T$) due to the MCE. In this method, the heat-capacity setup is directly employed, enabling simultaneous measurement of both heat capacity and the Gr\"uneissen parameter. The variations of $\Gamma_m$ vs. $B$ for three temperatures are shown in Fig.~\ref{FigS10}.

\section{Pressure-dependent Magnetization}
Magnetization studies under pressure were performed in the QD MPMS SQUID magnetometer. A single crystal of BCAO was loaded into the CuBe cell. Pressure was determined by measuring the superconducting transition of a small piece of Pb. Daphne oil was used as a pressure transmitting medium. The details of the experimental procedure can be found in~\cite{Shen2021}.
\section{Single Crystal Neutron Diffraction}
Single-crystal neutron diffraction was performed on the neutron diffractometer ZEBRA at SINQ, PSI. The crystal of the size $3\times 3\times 0.3$ mm$^3$ was mounted into the 10 T vertical magnet with the $b$-axis vertical. Neutrons of the wavelength 1.383~$\text{Å}$ selected by the Ge-monochromator were used in the normal beam geometry.
\clearpage

\begin{figure}
	\includegraphics[width= 14 cm]{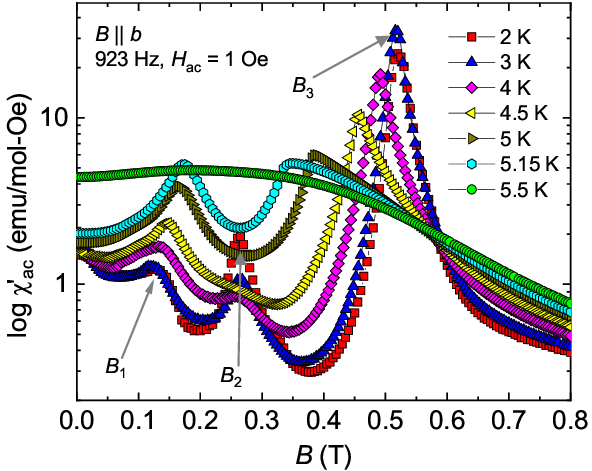}
	\caption{Magnetic field dependence of $\chi_{ac}^{\prime}$ for 5.5~K $\leq T \leq 2$~K for $B \parallel b$. Three clear transitions marked as $B_1$, $B_2$, and $B_3$ are observed at 2~K. As temperature increases up to $T_N$: $B_1$ initially remains constant, then steadily increases, $B_2$ remains stable until 4~K, then abruptly disappears above this, $B_3$ exhibits a consistent decrease beyond 3~K.  
		\label{FigS3}}
\end{figure}

\begin{figure}
	\includegraphics[width= 14 cm]{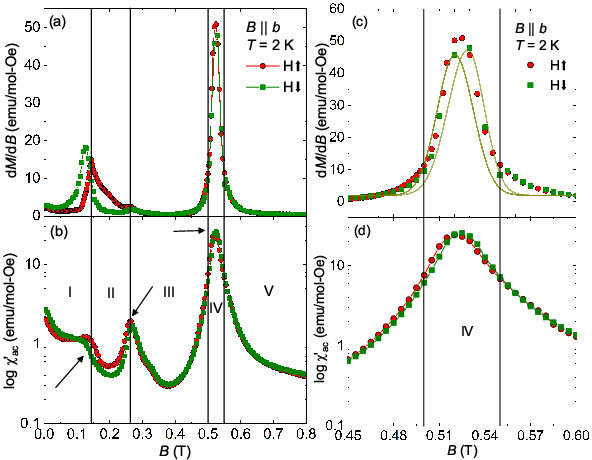}
	\caption{(a,b) Magnetic field dependence of differential susceptibility ($dM/dB$) and $\chi_{ac}^{\prime}$ for both up and down sweep cycles at $2$~K. The arrow marks indicates positions of $B_1$, $B_2$, and $B_3$ similar to Fig.~\ref{FigS3}. (c,d) Enlarged plots of $\chi_{ac}^{\prime}$ and $dM/dB$ in phase IV regime. The yellow solid lines are visual guides illustrating the two transitions in $dM/dB$. {Both $dM/dB$ and $\chi_{ac}^{\prime}$show the similar sequence of field-induced phase transitions, but with a different shape of the transition anomalies, probably because $\chi_{ac}^{\prime}$ includes dynamic effects, whereas the measurement of $dM/dB$ is static. Note that the double nature of the transition at $0.50-0.54$\,T is best seen in the $dM/dB$ data.}
		\label{FigS4}}
\end{figure}

\begin{figure}
	\includegraphics[width= \linewidth]{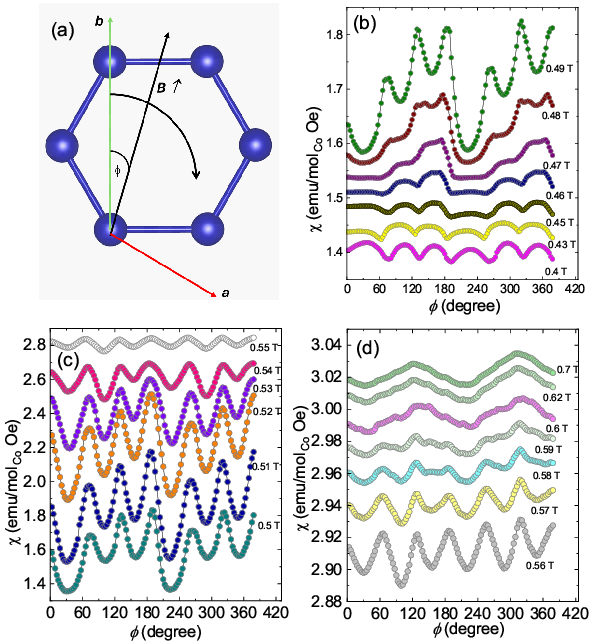}
	\caption{(a) Definition of the rotation angle $\phi$ used for the angle-dependent susceptibility. (b)-(d) Linear plots of angle-dependent magnetic susceptibility collected at $T$ =2.4\,K for different applied fields. Deviations from the 6-fold symmetry appear above 0.45\,T and culminate in the formation of phase IV that shows the same angular dependence across its stability range between 0.50 and 0.54\,T. Note that the maxima and minima of the magnetic susceptibility are swapped in phase IV compared to phase III, thus further proving the distinct nature of phase IV.The absolute values of the susceptibility are offset for clarity.
		\label{FigS5}}
\end{figure}
\begin{figure}
	\includegraphics[width= 12 cm]{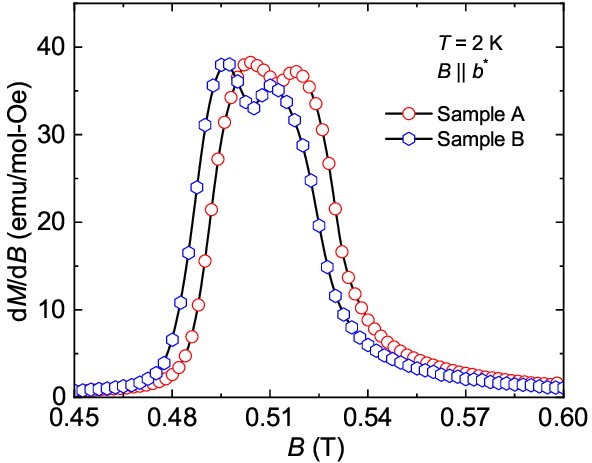}
	\caption{Sample dependency of double peak in $dM/dB$ vs. $B$ plots for phase IV.
		\label{FigS6}}
\end{figure}
\begin{figure}[htb]
	\includegraphics[width= 12 cm]{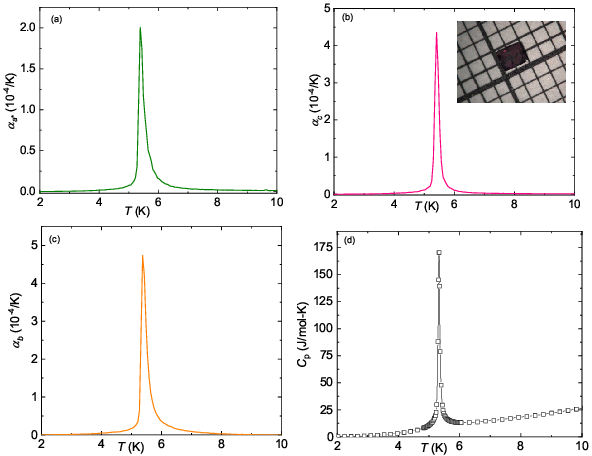}
	\caption{(a,b,c) Plots of the linear TE coefficient $\alpha$ vs $T$ for the three crystallographic directions. (d) Zero-field specific heat vs $T$ for 2~K $\leq T \leq$ 10~K. Note that the peak in panel (b) has a lower magnitude compared to Fig. 3b of the main 
		text. This is because the data were collected on a different crystal with the broadened transition anomaly.}
\end{figure}

\begin{figure}
	\includegraphics[width= 12 cm]{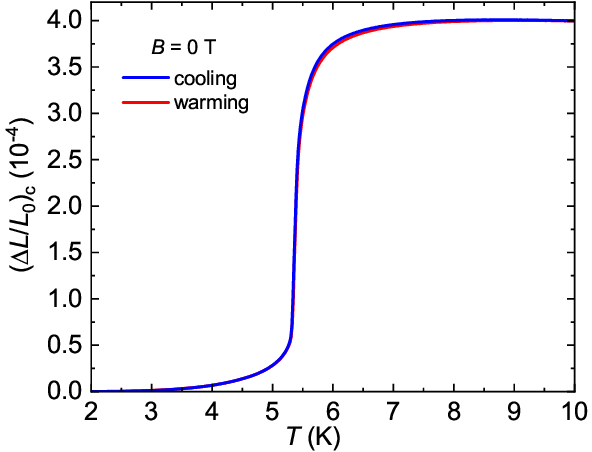}
	\caption{Relative length change measured at zero magnetic field on cooling and warming shows reversibility and indicates second-order nature of the phase transition in zero field. No anomalies were observed above 10~K up to 300~K.
		\label{FigS8}}
\end{figure}

\begin{figure}[htb]
	\includegraphics[width= \linewidth]{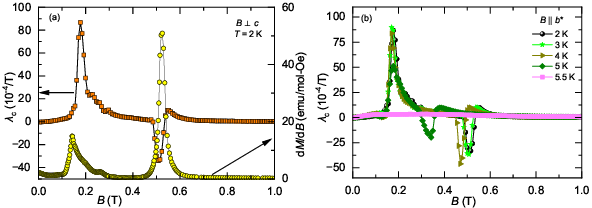}
	\caption{(a) Magnetic field dependence of $\lambda_{c}$ and $dM/dB$ plotted in left and right $y$-axes respectively at $T = 2$~K for the sweep up field cycle. (b) Plots of $\lambda_{c}$ vs. $B$ measured up to $1$~T for 5.5~K $\leq T \leq 2$~K.
		\label{FigS9}}
\end{figure}
\begin{figure}
	\includegraphics[width= 12 cm]{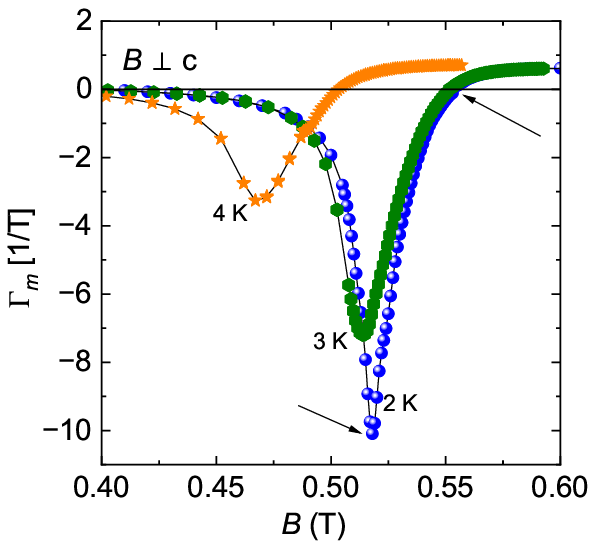}
	\caption{Magnetic field dependence of the magnetic Gr\"uneissen parameter in phase IV. The two transitions associated with phase IV can be identified from the peak and zero crossing of the $\Gamma_m$ plots at each temperatures.
		\label{FigS10}}
\end{figure}

\begin{figure}
	\includegraphics[width= 10 cm]{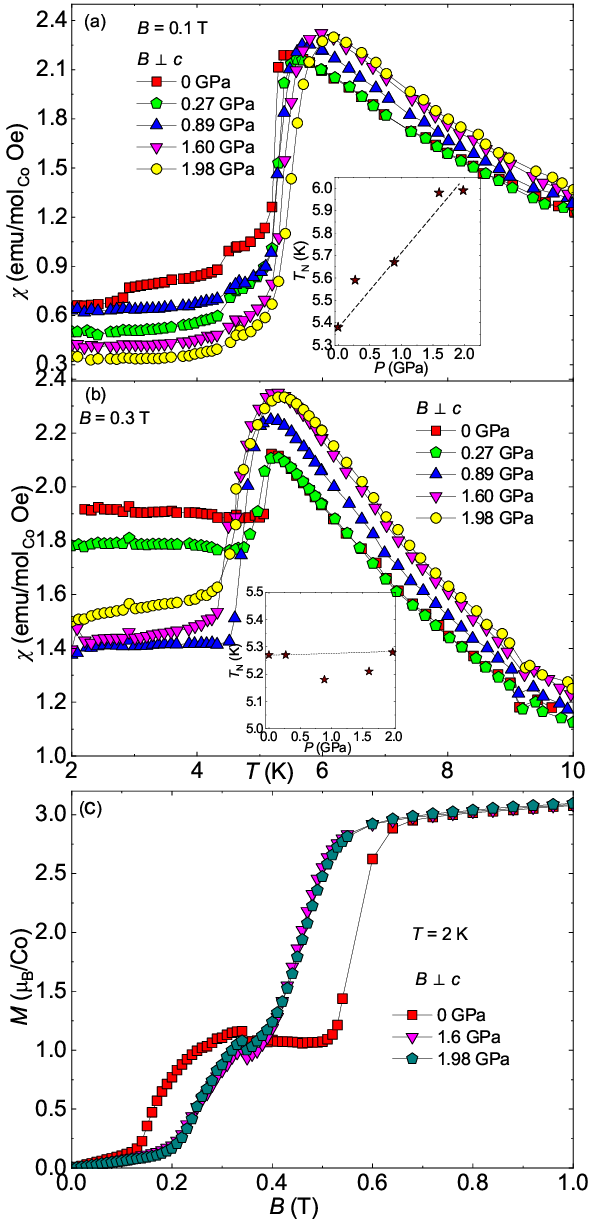}
	\caption{(a,b) Temperature-dependent DC magnetic susceptibility $\chi(T)$ measured under different pressures from 2 to 10~K under 0.1 and 0.3~T fields. Inset: Corresponding pressure--temperature phase diagram. (c) Magnetic isotherms measured up to 1~T at 2~K for different applied pressures.
		\label{FigS11}}
\end{figure}

\begin{figure}
	\includegraphics[width=\linewidth]{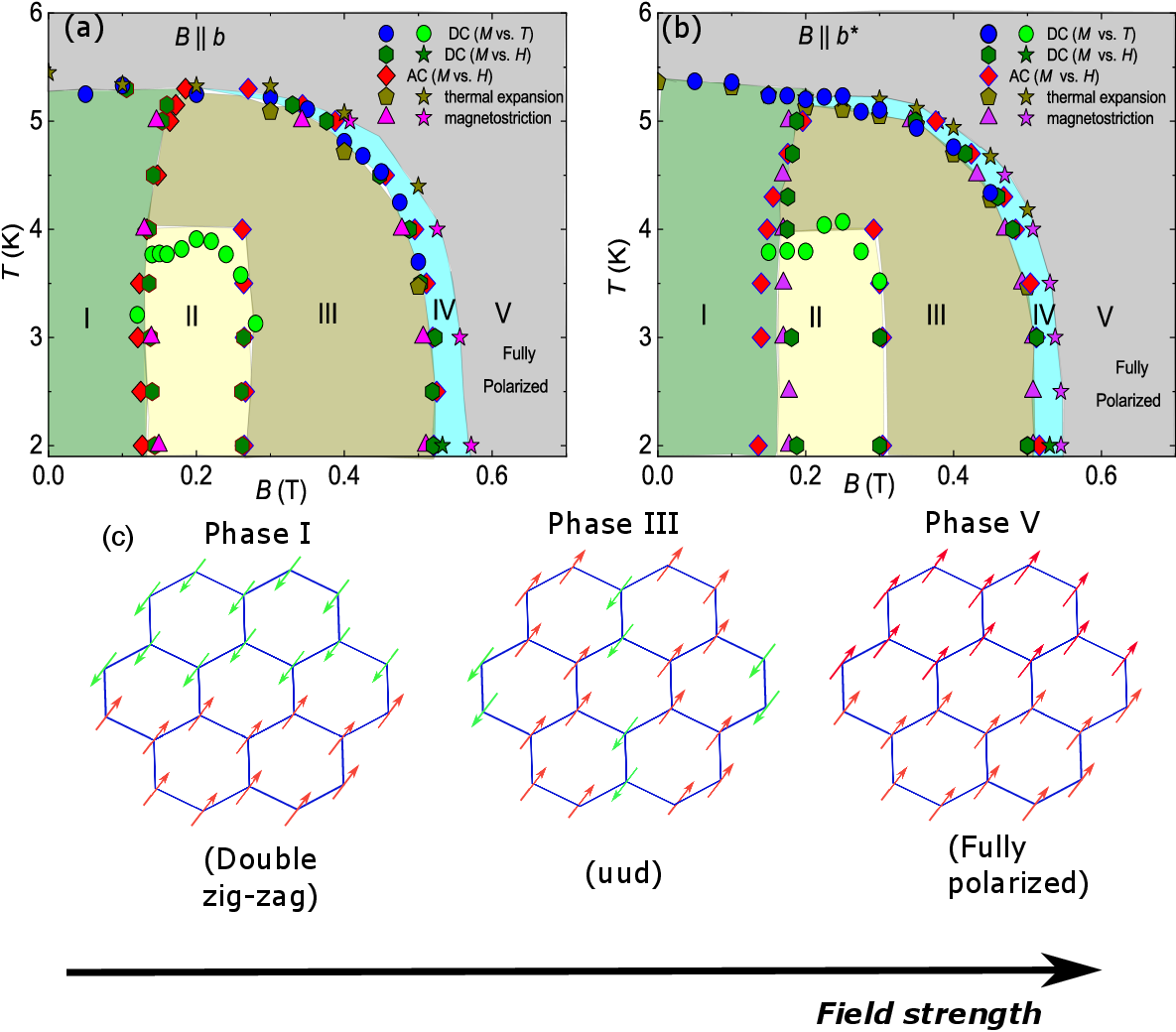}
	\caption{ (a,b) Field-temperature phase diagrams of BCAO constructed from the magnetometry and dilatometry measurements, respectively. (c) Schematic presentation of phases I, III, and V in BCAO.
		\label{FigS12}}
\end{figure}
\end{document}